\documentclass[letterpaper, conference, 10 pt]{ieeeconf}
\IEEEoverridecommandlockouts
\usepackage[english]{babel}

\let\labelindent\relax

\usepackage{enumitem}

\usepackage{amsmath, amssymb, amsthm}
\usepackage{mathtools}

\usepackage{arydshln}
\usepackage{booktabs,caption}
\usepackage[flushleft]{threeparttable}
\usepackage{graphicx}
\usepackage{subcaption}
\usepackage{caption}
\usepackage{multirow}
\usepackage{tabularx}
\usepackage{url}
\usepackage{accents}
\usepackage{cite}

\graphicspath{ {figures/} }
\usepackage{xcolor}
\usepackage{caption}
\usepackage{float}
\usepackage[linesnumbered,ruled,vlined,algo2e]{algorithm2e}
\usepackage{algorithmic}

\newtheorem{lemma}{Lemma}
\newtheorem{remark}{Remark}
\newtheorem{theorem}{Theorem}
\newtheorem{assumption}{Assumption}

\DeclarePairedDelimiter{\norm}{\lVert}{\rVert}

\SetKwInOut{Initial}{Initial}
\SetKwInOut{Parameter}{Parameters}
\SetKw{Continue}{continue}
\allowdisplaybreaks

\newcommand{\ubar}[1]{\underaccent{\bar}{#1}}

\def\BibTeX{{\rm B\kern-.05em{\sc i\kern-.025em b}\kern-.08em
    T\kern-.1667em\lower.7ex\hbox{E}\kern-.125emX}}

\title{\LARGE \bf
A Data-Driven Approach for Inverse Optimal Control
}

\author{Zihao Liang, \ Wenjian Hao, \ Shaoshuai Mou
\thanks{The authors are with the School of Aeronautics and Astronautics, Purdue University, IN 47907, USA {\tt\small \{liang331, hao93, mous\}@purdue.edu}}
}

\begin{document}

\maketitle
\thispagestyle{empty}
\pagestyle{empty}


\begin{abstract}
 This paper proposes a data-driven, iterative approach for inverse optimal control (IOC), which aims to learn the objective function of a nonlinear optimal control system given its states and inputs. The approach solves the IOC problem in a challenging situation when the system dynamics is unknown. The key idea of the proposed approach comes from the deep Koopman representation of the unknown system, which employs a deep neural network to represent observables for the Koopman operator. By assuming the objective function to be learned is parameterized as a linear combination of features with unknown weights, the proposed approach for IOC is able to achieve a Koopman representation of the unknown dynamics and the unknown weights in objective function together. Simulation is provided to verify the proposed approach.
\end{abstract}

\section{Introduction}


As one of the key techniques developed in control theories, \emph{optimal control} aims to find control inputs for a target system such that inputs and systems' states optimize a known objective function, which usually represents various mission objectives in practical applications. Such objective functions are usually unknown especially for complicated and newly developed missions such as human motion analysis \cite{WDJSS19TRO}, manipulation \cite{englert2017inverse}, human-robot interaction \cite{mainprice2016goal}, and autonomous driving \cite{kuderer2015learning}, for which limited knowledge is available and objective functions are usually implicit. To address this, researchers have recently devoted a large amount of attentions to \textbf{\emph{Inverse Optimal Control} (IOC)}, which aims to learn the objective function from observations of an expert system's trajectories (namely, inputs and states).



Most IOC methods typically assume the unknown objective function parameterized as a linear combination of selected prescribed features (or basis functions), where  each feature characterizes one aspect of the system behavior, such as energy cost, time consumption, risk levels, etc. The problem of solving IOC is changed to estimate the unknown weights in constructing the objective function based on such selected features. A direction to solving IOC problems is by adopting a double-layer architecture \cite{ng2000algorithms,mombaur2010human,abbeel2004apprenticeship,kuderer2015learning,mainprice2016goal,ratliff2006maximum,ziebart2008maximum}, in which the weights are updated in an outer layer while optimal control systems are solved in the inner layer with the cost of high computation for repeatedly solving optimal control problems. To further reduce the computational burden in solving IOC, researchers started to leverage the optimality conditions such as Karush-Kuhn-Tucker (KKT) conditions, for which the observed trajectory must satisfy, and the unknown weights can thus be solved directly solved by constructing the optimal equations \cite{keshavarz2011imputing,puydupin2012convex}. Along this direction, the authors of \cite{liang2022iterative, WDSS21IJRR} have solved the problem of IOC in the case when observed trajectories are not complete, with a further generalization of the proposed approach to a distributed algorithm for IOC in multi-agent systems in \cite{jin2021distributed}.

Note that all the IOC methods mentioned above heavily depend on the exact knowledge of the underlying dynamics of an optimal control system, i.e. these methods are not applicable if the dynamics are unknown. Obtaining a dynamics model is sometimes effort-demanding especially for high-dimensional systems as it requires a large amount of expertise and knowledge in systems and their motion \cite{schon2011system,nelles2001nonlinear}. This requirement in turn weakens one of the most  prominent benefits of  IOC techniques, which claims to empower  non-expert users to program the robot without much effort and only by   providing demonstrations. Recognition of this has motivated the goal of this paper, which aims to solve the IOC  problem even when the exact knowledge of system dynamics is not available. This requires us to develop a method that not only learns the control objective function but also the dynamics model from  demonstration data as well. We note that the Koopman operator has recently been attractive in representing an unknown nonlinear system by a linear time-varying system \cite{williams2015data,budivsic2012applied,williams2016extending,proctor2018generalizing}, based on which controllers could be designed \cite{korda2018linear}. System identification based on the Koopman operator relies on carefully selecting the observables, for which the introduction of deep neural networks (DNN) has recently proved to be helpful \cite{hao2020DLKR,hao2022deep}.

\begin{figure}[h]
    \centering
    \includegraphics[width=0.45\textwidth]{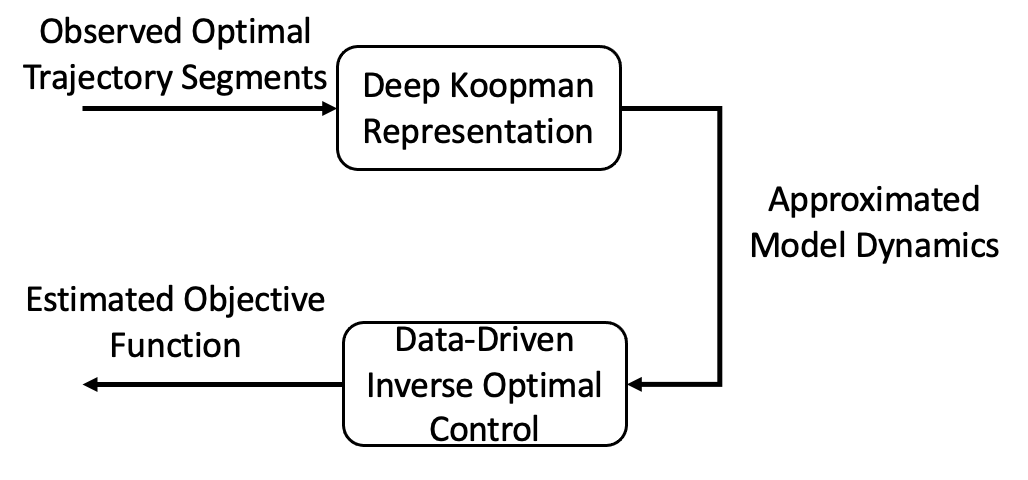}
    \caption{Data-Driven Inverse Optimal Control}
    \label{fig:DDIOC_flow}
\end{figure}

Motivated by the aforementioned limitation of existing IOC methods and the recent progress in applying Koopman-operator theory in solving data-driven control problems, this paper  develops a data-driven IOC approach, where jointly learn the unknown objective function and the underlying dynamics together. We first represent the unknown dynamics of an optimal control system using the Koopman operator, and then iteratively learn   the Koopman operator and the control objective function in the same learning framework. Figure \ref{fig:DDIOC_flow} illustrates the arrangement of the framework. Compared to existing IOC techniques, the proposed method  does not require information on system dynamics. Furthermore, the method does not necessarily require complete  demonstration data of an optimal control system, and the input data is allowed to be segments of optimal trajectories.


\noindent{\emph{Notations}}. Let $\parallel \cdot \parallel$ denote the Euclidean norm. For a matrix $A\in\mathbb{R}^{n\times m}$, $A^\prime$ denotes its transpose; $A^\dagger$ denotes its Moore-Penrose pseudoinverse. Let $\frac{\partial \boldsymbol{g}}{\partial \boldsymbol{x}_t}$ denotes the Jacobian matrix of a differentiable vector-valued function $\boldsymbol{g}(\boldsymbol{x})$ with respect to $\boldsymbol{x}$ evaluated at $\boldsymbol{x}_t$.

\section{Problem Formulation}
Consider a discrete-time optimal control system with the dynamics
\begin{equation}\label{eq:dyn_unknown}
\boldsymbol{x}_{t+1}=\boldsymbol{f}(\boldsymbol{x}_t,\boldsymbol{u}_t),
\end{equation}
where $\boldsymbol{x}_t\in\mathbb{R}^n$ is the system state;  $\boldsymbol{u}_t\in\mathbb{R}^m$ is the control input; $t=0,1,\cdots,T$ is the time step; and $\boldsymbol{f}:\mathbb{R}^n\times\mathbb{R}^m\rightarrow\mathbb{R}^n$ is \emph{unknown} and assumed to be differentiable. 
The control objective function of  the optimal control system is considered to be a linear combination of known features and \emph{unknown} weights: 
\begin{equation}\label{eq:objective_unknown}
J(\boldsymbol{x}_{0:T}, \boldsymbol{u}_{0:T},\boldsymbol{
\omega})=\sum_{t=0}^{T}\boldsymbol{\omega}^\prime\boldsymbol{\phi}(\boldsymbol{x}_t, \boldsymbol{u}_t),
\end{equation}
where $\boldsymbol{\phi}(\boldsymbol{x}, \boldsymbol{u}):\mathbb{R}^n\times\mathbb{R}^m\rightarrow\mathbb{R}^r$ is a specified feature vector function and assumed to be differentiable; $\boldsymbol{\omega}\in\mathbb{R}^r$ is a vector of  weights, which are \emph{unknown}; and $T$ is the time horizon. 




Since the system considered here is an optimal control system, any  trajectory of time horizon $T$, denoted as a sequence of states-inputs $\boldsymbol{\xi}_{0:T}=\{\boldsymbol{x}_{0:T},\boldsymbol{u}_{0:T}\}$, minimizes the cost function \eqref{eq:objective_unknown} and satisfies the dynamics (\ref{eq:dyn_unknown}).
Suppose a set of observed sequences of states-inputs is given, which is  denoted by
\begin{equation}\label{eq:data}
\mathcal{D}=\{\boldsymbol{\zeta}_1,\boldsymbol{\zeta}_2,\cdots,\boldsymbol{\zeta}_D  \}
\end{equation}
with each sequence being a segment of a system trajectory that minimizes the cost function \eqref{eq:objective_unknown}:
\begin{equation}\label{eq:sequence}
\boldsymbol{\zeta}_i=\{\boldsymbol{x}^*_{\ubar{t}_i:\bar{t}_i},\boldsymbol{u}^*_{\ubar{t}_i:\bar{t}_i}\}\subseteq\boldsymbol{\xi}_{0:T},
\end{equation}
with $\ubar{t}_i, \bar{t}_i$ being the starting time and end time of $i$th state-input sequence. 

The \textbf{goal} of this paper is to develop an algorithm to estimate the unknown weight vector $\boldsymbol{\omega}$ in \eqref{eq:objective_unknown} via IOC with the given dataset $\mathcal{D}$ without knowing the system dynamics.

\section{Main Results}

This section develops the data-driven inverse optimal control (IOC) algorithm. Here, we employ the deep Koopman representation (DKR) to approximate the unknown dynamics (\ref{eq:dyn_unknown}). The first part below presents the deep Koopman representation of unknown dynamics, the second part presents the IOC method based on Koopman operator dynamics, and the third part develops the data-driven IOC algorithm, where the objective function and the Koopman operator dynamics are jointly learned.

\subsection{Dynamics Approximation using Deep Koopman Representation}\label{section:dynamics}
In this section, we focus on the data-driven approximation of unknown system dynamics. We employ DKR as in \cite{hao2020DLKR}. DKR uses the nonlinear mapping $\boldsymbol{\psi}(\cdot,\boldsymbol{\theta}):\mathbb{R}^n\rightarrow\mathbb{R}^N$, parameterized by $\boldsymbol{\theta}\in\mathbb{R}^q$, as the finite-dimension Koopman observable. $\boldsymbol{\psi}(\cdot,\boldsymbol{\theta})$ is represented by a Deep Neural Network (DNN) with a known structure but an unknown parameter $\boldsymbol{\theta}$ to be determined by the set of observed trajectories $\mathcal{D}$. We also denote the number of hidden layers nodes as $n_h$. One can approximate the unknown dynamics \eqref{eq:dyn_unknown} by finding $\boldsymbol{\psi}(\cdot,\boldsymbol{\theta})$ and matrices  $\mathcal{K}_x\in\mathbb{R}^{N\times N}$, $\mathcal{K}_u\in\mathbb{R}^{N\times m}$, $\mathcal{C}\in\mathbb{R}^{n\times N}$ based on dataset $\mathcal{D}$ such that for $t\le T$,
\begin{equation} \label{eq:obsevo}
\begin{aligned}
    \boldsymbol{\psi}(\boldsymbol{x}_{t+1},\boldsymbol{\theta})&=\mathcal{K}_x\boldsymbol{\psi}(\boldsymbol{x}_t,\boldsymbol{\theta})+\mathcal{K}_{{u}}\boldsymbol{u}_t,\\
    \hat{\boldsymbol{x}}_{t+1}&=\mathcal{C}\boldsymbol{\psi}(\boldsymbol{x}_{t+1},\boldsymbol{\theta}),
\end{aligned}
\end{equation}
where $\hat{\boldsymbol{x}}_t\in\mathbb{R}^n$ is the estimated states vector obtained by DKR.
By rewriting \eqref{eq:obsevo}, one can achieve:
\begin{equation}\label{eq:koopman_LTI}
\begin{aligned}
\hat{\boldsymbol{x}}_{t+1}&=\hat{\boldsymbol{f}}(\boldsymbol{x}_t,\boldsymbol{u}_t)\\
&= \mathcal{C}\boldsymbol{\psi}(\boldsymbol{x}_{t+1},\boldsymbol{\theta})\\
&= 
\mathcal{C}\mathcal{K}_x\boldsymbol{\psi}(\boldsymbol{x}_t,\boldsymbol{\theta})+\mathcal{C}\mathcal{K}_{{u}}\boldsymbol{u}_t,
\end{aligned}
\end{equation}
where $\hat{\boldsymbol{f}}$ denotes the approximated system of \eqref{eq:dyn_unknown}. 

It is noted from (\ref{eq:koopman_LTI}) that the Koopman operator for approximation of the dynamical system is to transfer the system dynamics (\ref{eq:dyn_unknown}) into a linear system which has the observables $\boldsymbol{\psi}(\boldsymbol{x}_t,\boldsymbol{\theta})$ as its state.  This linear system facilitates the analysis of the original non-linear control system, especially in the field of system learning \cite{abraham2019active}, nonlinear control \cite{korda2018linear}, etc.

We define a vector $\boldsymbol{z}(\boldsymbol{x}_t,\boldsymbol{u}_t,\boldsymbol{\theta})\in \mathbb{R}^{N+m}$ consists of the observables $\boldsymbol{\psi}(\boldsymbol{x}_t,\boldsymbol{\theta})\in\mathbb{R}^{N}$ over states  (normally $N\gg n$), and the inputs $\boldsymbol{u}_t$:
\begin{equation}\label{eq:zobservable}
\boldsymbol{z}(\boldsymbol{x}_t,\boldsymbol{u}_t,\boldsymbol{\theta})=\begin{bmatrix}
\boldsymbol{\psi}(\boldsymbol{x}_t,\boldsymbol{\theta})\\
\boldsymbol{u}_t
\end{bmatrix}.
\end{equation}

Then, the finite-dimensional Koopman operator $\mathcal{K}:\mathbb{R}^{N+m}\rightarrow\mathbb{R}^{N+m}$ that acts on the space spanned by all  observables in $\boldsymbol{z}(\boldsymbol{x}_t,\boldsymbol{u}_t,\boldsymbol{\theta})$ can be written as:
\begin{equation}\label{eq:koopman-K-inputs}
\mathcal{K}\eqqcolon\begin{bmatrix}
\mathcal{K}_x& \mathcal{K}_u\\
\end{bmatrix}\in\mathbb{R}^{N\times (N+m)}.
\end{equation} 

Thus, by combining \eqref{eq:koopman_LTI}-\eqref{eq:koopman-K-inputs}, with a given pair of states-inputs $\{
\boldsymbol{x}_t,
\boldsymbol{u}_t\}$, the DKR approximation of dynamical system (\ref{eq:dyn_unknown}) is  
\begin{equation}\label{eq:approximate_dyn}
\begin{aligned}
\hat{\boldsymbol{x}}_{t+1}
&=\mathcal{C}\mathcal{K}_x\boldsymbol{\psi}(\boldsymbol{x}_t,\boldsymbol{\theta})+\mathcal{C}\mathcal{K}_{u}\boldsymbol{u}_t\\
&=C\mathcal{K}\boldsymbol{z}(\boldsymbol{x}_t,\boldsymbol{u}_t,\boldsymbol{\theta}).
\end{aligned}
\end{equation}


For any  sequence $\boldsymbol{\zeta}_i\in\mathcal{D}$, we can define the following dynamics approximation loss:
\begin{equation}\label{eq:koopmanloss}
l_{\mathcal{K}}^i(\mathcal{K},\boldsymbol{\zeta}_i,\boldsymbol{\theta})=\frac{1}{\tau}\sum_{t=\ubar{t}_i}^{\bar{t}_i-1}\norm{\boldsymbol{\psi}(\boldsymbol{x}_{t+1},\boldsymbol{\theta})- \mathcal{K}\boldsymbol{z}(\boldsymbol{x}_t,\boldsymbol{u}_t,\boldsymbol{\theta})}^2,
\end{equation}
where $\tau =\bar{t}_{i}-\ubar{t}_{i}$.
 The $\mathcal{C}$ matrix is computed by minimizing the following loss function:
\begin{equation}\label{eq:compute-C}
l_{\mathcal{C}}^i=\frac{1}{\tau}\sum_{t=\ubar{t}_i}^{\bar{t}_i-1}\norm{\boldsymbol{x}_{t}-\mathcal{C}\boldsymbol{\psi}(\boldsymbol{x}_{t},\boldsymbol{\theta})}^2.
\end{equation}



To solve (\ref{eq:koopmanloss}) analytically, we first define:
\begin{equation}\label{eq:PSI_x}
    \boldsymbol{\Psi}_i^x = [\boldsymbol{\psi}(\boldsymbol{x}_{\ubar{t}_i},\boldsymbol{\theta}),...,\boldsymbol{\psi}(\boldsymbol{x}_{\bar{t}_i-1},\boldsymbol{\theta})],
\end{equation}
\begin{equation}\label{eq:PSI_(x+1)}
    \boldsymbol{\Psi}_i^{x+1} =[\boldsymbol{\psi}(\boldsymbol{x}_{\ubar{t}_i+1},\boldsymbol{\theta}),...,\boldsymbol{\psi}(\boldsymbol{x}_{\bar{t}_i},\boldsymbol{\theta})],
\end{equation}
\begin{equation}
    \boldsymbol{U}_i =[\boldsymbol{u}_{\ubar{t}_i},...,\boldsymbol{u}_{\bar{t}_i-1}],
\end{equation}
\begin{equation} \label{eq:Z_i}
    \boldsymbol{Z}_i =
    \begin{bmatrix}
         \boldsymbol{\Psi}_i^x\\
         \boldsymbol{U}_i
    \end{bmatrix}
    .
\end{equation}

The Koopman operator is computed analytically by solving:


\begin{equation} \label{eq:compute-K}
    \mathcal{K}=\boldsymbol{\Psi}_i^{x+1}
        \boldsymbol{Z}_i^\prime
    (  \boldsymbol{Z}_i
    \boldsymbol{Z}_i^\prime)^{-1}.
\end{equation}
Any solution to (\ref{eq:compute-K}) is a solution to (\ref{eq:koopmanloss}) \cite{korda2018linear}. 
We are also able to solve the equation \eqref{eq:compute-C} analytically for the matrix $\mathcal{C}$ by:
\begin{equation}\label{eq:compute-C-analytical}
    \mathcal{C}=X_i(\boldsymbol{\Psi}_i^x)^{\dagger}
\end{equation}
where $X_i=[\boldsymbol{x}_{\ubar{t}_{i}},\cdots,\boldsymbol{x}_{\bar{t}_{i}-1}]$.

Equation \eqref{eq:compute-K} computes the Koopman operator by only utilizing a segment of trajectory $\boldsymbol{\zeta}_i$ in the provided set of observed data $\mathcal{D}$. To incorporate other segments of the trajectory, one needs to compute the inverse in (\ref{eq:compute-K}) and the pseudo-inverse in (\ref{eq:compute-C-analytical}) repeatedly, which is computationally expensive as $i$ increases. To fully utilize the whole data set in a computationally efficient way, we borrow the iterative update law of the Koopman operator proposed by \cite{hao2022deep}. To utilize this update law, the following assumptions need to be made:

\begin{assumption} \label{assumption:fullrank}
The matrices $\boldsymbol{\Psi}_i^x$ in \eqref{eq:PSI_x} and $\boldsymbol{Z}_i$ in \eqref{eq:Z_i} are of full row rank.

 \end{assumption}

\begin{remark}
Assumption \ref{assumption:fullrank} ensures that the matrices $\boldsymbol{\Psi}_i^x$ and $\boldsymbol{Z}_i$ are invertible.
\end{remark}

\begin{assumption} \label{assumption:deltaT}
For any $\boldsymbol{\zeta}_i$ in \eqref{eq:data}, let $\Delta t$ denotes the observation interval between each states-inputs pair $\{\boldsymbol{x}_t,\boldsymbol{u}_t\}$. The observation interval $\Delta t$ is sufficiently small such that for some constant $\mu_x \geq 0, \mu_u\geq 0$, $\parallel \boldsymbol{x}_{t+1} - \boldsymbol{x}_{t} \parallel < \mu_x < \infty$ and $\parallel \boldsymbol{u}_{t+1} - \boldsymbol{u}_t \parallel < \mu_u < \infty$.
 \end{assumption}

\begin{remark}
If the observation interval $\Delta t$ goes to zero, the constant $\mu_x$ and $\mu_u$ also go to zero.
\end{remark}

\begin{assumption}\label{assumption:lipschitz}
    The deep neural network observable function $\boldsymbol{\psi}(\boldsymbol{x},\boldsymbol{\theta})$ is Lipschitz continuous on the system state space with Lipschitz constant $\mu_g$.
\end{assumption}

With assumptions \ref{assumption:fullrank}-\ref{assumption:lipschitz} are made, the lemma of the update law is introduced:

\begin{lemma}\label{lemma:KoopmanUpdate}
\cite{hao2022deep} If assumption \ref{assumption:fullrank}-\ref{assumption:lipschitz} hold, given $\boldsymbol{\psi}(\cdot,\boldsymbol{\theta})$, $\mathcal{K}$ and $\mathcal{C}$, with a new batch of data denoted as $\boldsymbol{\zeta}_{i+1}$, the Koopman operator $\mathcal{K}$ and matrix $\mathcal{C}$ are updated as follows:
\begin{equation} \label{eq:update-K}
    \mathcal{K}=(\boldsymbol{\Psi}_{i+1}^{x+1}-\mathcal{K}\boldsymbol{Z}_{i+1})\gamma_{i+1}\boldsymbol{Z}^{\prime}_{i+1}(\boldsymbol{Z}_{i}\boldsymbol{Z}_{i}^\prime)^{-1}+\mathcal{K},
\end{equation}
{\small 
\begin{equation} \label{eq:update-C}
    \mathcal{C}=(\boldsymbol{Z}_{i+1}-\mathcal{C}\boldsymbol{\Psi}_{i+1}^{x})\bar{\gamma}_{i+1}(\boldsymbol{\Psi}_{i+1}^{x})^\prime(\boldsymbol{\Psi}_i^{x}(\boldsymbol{\Psi}_i^{x})^\prime)^{-1}+\mathcal{C},
\end{equation}
}
where
\begin{equation}
    \gamma_{i+1}=(I_{\tau}+\boldsymbol{Z}^{\prime}_{i+1} (\boldsymbol{Z}_{i}\boldsymbol{Z}_{i}^\prime)^{-1}\boldsymbol{Z}_{i+1})^{-1}  \in \mathbb{R}^{\tau \times \tau},
\end{equation}
{\small
\begin{equation}
\bar{\gamma}_{i+1}=(I_{\tau}+(\boldsymbol{\Psi}_{i+1}^{x})^\prime (\boldsymbol{\Psi}_{i}^{x}(\boldsymbol{\Psi}_{i}^{x})^\prime)^{-1}\boldsymbol{\Psi}_{i+1}^{x})^{-1}  \in \mathbb{R}^{\tau \times \tau}.
\end{equation}

}
\end{lemma}

Once the matrices $\mathcal{K}$ and $\mathcal{C}$ are updated, the parameter $\boldsymbol{\theta}$ is solved by the following optimization problem:
\begin{equation}\label{eq:total_loss}
\boldsymbol{\theta}^*=\arg\min_{\boldsymbol{\theta}}\sum_{j=1}^{i+1} l_{\mathcal{K}}^j+l_{\mathcal{C}}^j.
\end{equation}

\subsection{Inverse Optimal Control with Deep Koopman Representation}

We consider a system trajectory of time horizon $T$, $\boldsymbol{\xi}_{0:T}=\{\boldsymbol{x}_{0:T},\boldsymbol{u}_{0:T}\}$, which minimizes the cost function given in (\ref{eq:objective_unknown}).  Based on Pontryagin's maximum principle \cite{pontryagin1962mathematical}, there exists a sequence of costates $\boldsymbol{\lambda}_{t}\in\mathbb{R}^n$ with $t=0,\cdots,T$, such that the following optimality conditions are satisfied:
\begin{equation}\label{pmp}
\begin{aligned}
\boldsymbol{\lambda}_t&=\frac{\partial \boldsymbol{\phi}^\prime}{\partial \boldsymbol{x}_t}\boldsymbol{\omega}+\frac{\partial \boldsymbol{f}^\prime}{\partial \boldsymbol{x}_t}\boldsymbol{\lambda}_{t+1},\\
\boldsymbol{0}&=\frac{\partial \boldsymbol{\phi}^\prime}{\partial \boldsymbol{u}_t}\boldsymbol{\omega}+\frac{\partial \boldsymbol{f}^\prime}{\partial \boldsymbol{u}_t}\boldsymbol{\lambda}_{t+1},
\end{aligned}
\end{equation}
for $t=0,1,\cdots, T-1$, and $\boldsymbol{\lambda}_{T}=\frac{\partial \boldsymbol{\phi}^\prime}{\partial \boldsymbol{x}_T}\boldsymbol{\omega}$. 

Now, we replace $\boldsymbol{f}$ with $\hat{\boldsymbol{f}}$ that is obtained using DKR. According to \eqref{eq:approximate_dyn}, we have:
\begin{equation} \label{eq:approx_dfdx}
\begin{aligned}
    \frac{\partial\hat{\boldsymbol{f}}}{\partial\boldsymbol{x}_t}&=\mathcal{C}\mathcal{K}_x\frac{\partial\boldsymbol{\psi}(\boldsymbol{x}_t,\boldsymbol{\theta})}{\partial\boldsymbol{x}_t},\\
    \frac{\partial\hat{\boldsymbol{f}}}{\partial\boldsymbol{u}_t}&=\mathcal{C}\mathcal{K}_u.
\end{aligned}
\end{equation}
Now, we substitute \eqref{eq:approx_dfdx} into (\ref{pmp})  leads to 
\begin{equation}\label{pmp2}
\begin{aligned}
\boldsymbol{\lambda}_t&=\frac{\partial \boldsymbol{\phi}^\prime}{\partial \boldsymbol{x}_t}\boldsymbol{\omega}+\frac{\partial \boldsymbol{\psi}^\prime}{\partial \boldsymbol{x}_t}\mathcal{K}_x^\prime C^\prime\boldsymbol{\lambda}_{t+1},\\
\boldsymbol{0}&=\frac{\partial \boldsymbol{\phi}^\prime}{\partial \boldsymbol{u}_t}\boldsymbol{\omega}+ \mathcal{K}_u^\prime C^\prime\boldsymbol{\lambda}_{t+1}.
\end{aligned}
\end{equation}
Note that in \eqref{pmp2}, $\boldsymbol{\psi}(\boldsymbol{x}_t,\boldsymbol{\theta})$ is represented by $\boldsymbol{\psi}$ for simplicity.

Now, we consider a segment of the system trajectory data, say $\boldsymbol{\zeta}=\{\boldsymbol{x}_{\ubar{t}:\bar{t}}, \boldsymbol{u}_{\ubar{t}:\bar{t}} \}\subseteq\boldsymbol{\xi}_{0:T}$. By writing (\ref{pmp2}) in matrix form corresponding to the available data $\boldsymbol{\zeta}$, we have the following compact equation
\begin{equation}\label{PMP}
\begin{aligned}
\boldsymbol{A}\boldsymbol{\lambda}_{\ubar{t}+1:\bar{t}}-\boldsymbol{\Phi}_x\boldsymbol{\omega}&=\boldsymbol{V}\boldsymbol{\lambda}_{\bar{t}+1},\\
\boldsymbol{B}\boldsymbol{\lambda}_{\ubar{t}+1:\bar{t}}+\boldsymbol{\Phi}_u\boldsymbol{\omega}&=\boldsymbol{0},
\end{aligned}
\end{equation}
where 
\begin{align}
\boldsymbol{A}&=
\begin{bmatrix}
{I} & \frac{-\partial \boldsymbol{\psi}^\prime}{\partial \boldsymbol{x}_{\ubar{t}+1}} \scriptstyle{\mathcal{K}_x^\prime C^\prime} & \cdots  & 0 & 0\\
0   & I & \cdots & 0  & 0\\
\vdots   & \vdots & \ddots &  \ddots & \vdots \\
0 & 0  &  \cdots & I & \frac{-\partial \boldsymbol{\psi}^\prime}{\partial \boldsymbol{x}_{\bar{t}-1}}\scriptstyle{\mathcal{K}_x^\prime C^\prime} \\
0 & 0 & \cdots & 0 & I
\label{A}
\end{bmatrix},\\
\boldsymbol{B}&=\begin{bmatrix}
\scriptstyle{\mathcal{K}_u^\prime C^\prime} & 0  & \cdots  & 0 \\
0 & \scriptstyle{\mathcal{K}_u^\prime C^\prime} &  \cdots  & 0 \\
 \vdots &  &  \ddots & \vdots  \\
 0 & 0  & \cdots &\scriptstyle{\mathcal{K}_u^\prime C^\prime}
 \label{B}
\end{bmatrix},\\
\boldsymbol{\Phi}_x&=\begin{bmatrix}
\frac{\partial \boldsymbol{\phi}}{\partial \boldsymbol{x}_{\ubar{t}+1}}& \frac{\partial \boldsymbol{\phi}}{\partial \boldsymbol{x}_{\ubar{t}+2}} & \cdots & \frac{\partial \boldsymbol{\phi}}{\partial \boldsymbol{x}_{\bar{t}-1}} & \frac{\partial \boldsymbol{\phi}}{\partial \boldsymbol{x}_{\bar{t}}}
\end{bmatrix}^\prime,\\
\boldsymbol{\Phi}_u&=\begin{bmatrix}
\frac{\partial \boldsymbol{\phi}}{\partial \boldsymbol{u}_{\ubar{t}}}& \frac{\partial \boldsymbol{\phi}}{\partial \boldsymbol{u}_{\ubar{t}+1}} & \cdots & \frac{\partial \boldsymbol{\phi}}{\partial \boldsymbol{u}_{\bar{t}-2}} & \frac{\partial \boldsymbol{\phi}}{\partial \boldsymbol{u}_{\bar{t}-1}}
\end{bmatrix}^\prime,\\
\boldsymbol{V}&=\begin{bmatrix}
0 & 0  & \cdots & 0 & \scriptstyle{C\mathcal{K}_x}\frac{\partial \boldsymbol{\psi}}{\partial \boldsymbol{x}_{\bar{t}}}
\end{bmatrix}^\prime.
\end{align}

The equations in (\ref{PMP}) establishes the relationship between data and the unknown weight vector. It can be written as follow:
\begin{equation} \label{eq:Fw=0}
\underbrace{
\begin{bmatrix}
\boldsymbol{A}&-\boldsymbol{\Phi}_x&\boldsymbol{V}\\
\boldsymbol{B}&\boldsymbol{\Phi}_u&\boldsymbol{0}
\end{bmatrix}
}
_{\hat{\boldsymbol{F}}(\mathcal{K}, \mathcal{C}, \boldsymbol{\zeta})}
\underbrace{
\begin{bmatrix}
\boldsymbol{\lambda}_{\ubar{t}+1:\bar{t}}\\
\boldsymbol{\omega}\\
\boldsymbol{\lambda}_{\bar{t}+1}
\end{bmatrix}
}
_{\boldsymbol{\nu}(\boldsymbol{\lambda},\boldsymbol{\omega})}
=\boldsymbol{0}
\end{equation}
with $\hat{\boldsymbol{F}}\in\mathbb{R}^{(n+m)(\bar{t}-\ubar{t}+1)\times (n(\bar{t}-\ubar{t}+2)+r)}$ depends on the Koopman operator $\mathcal{K}$, matrix $\mathcal{C}$  and the data segment $\boldsymbol{\zeta}$; $\boldsymbol{\nu}\in\mathbb{R}^{n(\bar{t}-\ubar{t}+2)+r}$ depends on the costates $\boldsymbol{\lambda}$ (includes $\boldsymbol{\lambda}_{\ubar{t}+1:\bar{t}}$ and $\boldsymbol{\lambda}_{\bar{t}+1}$) and unknown weight vector $\boldsymbol{\omega}$. Note that in \eqref{eq:Fw=0}, the notations $\hat{\boldsymbol{F}}$ and $\boldsymbol{F}$ mean the matrices are generated with the approximated system $\hat{\boldsymbol{f}}$ and true system $\boldsymbol{f}$ in \eqref{eq:dyn_unknown} respectively.

For a segment of the system trajectory data $\boldsymbol{\zeta}$, if the Koopman operator $\mathcal{K}$ and matrix $\mathcal{C}$ are given, one can choose to obtain a least square estimate for the weights $\boldsymbol{\omega}$ by solving the following equivalent optimization,
\begin{equation}\label{eq:iocloss}
\hat{\boldsymbol{\nu}}(\hat{\boldsymbol{\lambda}},\hat{\boldsymbol{\omega}})=\arg\min_{\boldsymbol{\nu}}\boldsymbol{\nu}^\prime\hat{\boldsymbol{F}}^\prime\hat{\boldsymbol{F}}\boldsymbol{\nu}.
\end{equation}
Here, $\hat{\boldsymbol{\nu}}$, $\hat{\boldsymbol{\lambda}}$ and $\boldsymbol{\hat{\omega}}$ are called a least-square estimate to the vector $\boldsymbol{\nu}$, costates $\boldsymbol{\lambda}$ and unknown weights $\boldsymbol{\omega}$ respectively. 
Also note that to prevent obtaining the trivial solution, a normalization constraint is typically added to the weight variables, e.g., $\sum_{i=1}^{r}\omega_i=1$.

\subsection{Data-Driven IOC Algorithm}
We now develop the data-driven IOC framework to estimate the unknown objective weight with unknown dynamics using the given dataset $\mathcal{D}$ in (\ref{eq:data}). To solve for the data-driven IOC problem, we propose a method to update the parameter $\boldsymbol{\theta}$ and the unknown weight $\boldsymbol{\omega}$ iteratively. The pseudo code of the proposed method is demonstrated in Algorithm \ref{algorithm}.


\begin{algorithm2e}[h]
	\caption{Data-driven IOC Pseudo Code} \label{algorithm}
	\DontPrintSemicolon
	\SetKwInOut{Input}{Input}
	\SetKwInOut{Output}{Output}
	\SetKwInput{Initialize}{Initialize}

	\Input{$\mathcal{D}, \boldsymbol{\phi}$}
	\Output{$\hat{\boldsymbol{\omega}}$}
	\Initialize {$\boldsymbol{\psi}(\boldsymbol{x}_t,\boldsymbol{\theta})$ 
 {\small //$\ $ Build Koopman observables using DNN with initial guess of $\boldsymbol{\theta}$.}}
    Obtain $\mathcal{K}$ and $\mathcal{C}$ by solving \eqref{eq:compute-K} and \eqref{eq:compute-C-analytical} with $\boldsymbol{\zeta}_1$.\;
    \For {$i=2:D$}{
    Update $\mathcal{K}$ and $\mathcal{C}$ with $\boldsymbol{\zeta}_i$ using  \eqref{eq:update-K} and \eqref{eq:update-C}.\;
    Solve \eqref{eq:total_loss} to obtain $\boldsymbol{\theta}$.\;
    Generate matrix $\hat{\boldsymbol{F}}$ with obtained $\mathcal{K}$, $\mathcal{C}$, $\boldsymbol{\theta}$ and all of the incorporated trajectory segments $\boldsymbol{\zeta}_j$, where $j\le i$.\;
    Solve \eqref{eq:iocloss} to obtain the least-square estimate of weight vector $\hat{\boldsymbol{\omega}}$.
    }
\end{algorithm2e}

We now present our main theorem:
\begin{theorem} \label{theorem}
Given a set of observed sequences of states-inputs pair (\ref{eq:data}). With the unknown model dynamics (\ref{eq:dyn_unknown}) approximated by DKR, which has observables $\boldsymbol{\psi}({\boldsymbol{x}_t,\boldsymbol{\theta}})$ represented by DNN and parameterized by $\boldsymbol{\theta}$. By using Algorithm \ref{algorithm}, the least-square estimate $\hat{\boldsymbol{\omega}}$ of the unknown objective weight in (\ref{eq:objective_unknown}) converges to the true weight $\boldsymbol{\omega}$ if assumptions \ref{assumption:fullrank}-\ref{assumption:lipschitz} hold and the following conditions are fulfilled:
\begin{enumerate}
    \item There are infinite number of hidden layer nodes in the DNN, i.e. $n_h=\infty$.
    \item The observation interval $\Delta t$ is sufficiently small such that $\mu_x$ and $\mu_u$ equal to zero.
    \item $\max_{\boldsymbol{x}_t\in\mathcal{D}}\parallel \boldsymbol{x}_t - \mathcal{C} \boldsymbol{\psi}(\boldsymbol{x}_t,\boldsymbol{\theta}) \parallel$ is zero.
\end{enumerate}
\end{theorem}

The proof of convergence of Algorithm \ref{algorithm} will be shown in the next section.

\subsection{Convergence Analysis}
This section provides the convergence analysis for the proposed data-driven IOC algorithm shown in Algorithm \ref{algorithm}. First, we denote the estimation error of the approximated system $\hat{\boldsymbol{f}}$ as $\boldsymbol{e}_t\in\mathbb{R}^n$, where
\begin{equation}
    \boldsymbol{e}_t = \hat{\boldsymbol{f}}(\boldsymbol{x}_t,\boldsymbol{u}_t)-\boldsymbol{f}(\boldsymbol{x}_t,\boldsymbol{u}_t).
\end{equation}
By using the update rule stated in Lemma~\ref{lemma:estimationError}, the norm of the estimation error $\boldsymbol{e}_t$ is bounded and can become zero according to the following lemma:

\begin{lemma} \label{lemma:estimationError}
    \cite{hao2022deep} If assumptions \ref{assumption:fullrank}-\ref{assumption:lipschitz} hold, then the supremum of the norm of the estimation error $\boldsymbol{e}_t$ is:
    \begin{equation}
        \begin{aligned}
        \lim_{n_h\rightarrow\infty} \sup\parallel \boldsymbol{e}_t\parallel =  (\parallel \mathcal{C} \mathcal{K}_x \parallel \mu_g + 1) \mu_x + \parallel \mathcal{C} \mathcal{K}_u \parallel  \mu_u \\+ l_{\mathcal{C}}^{max}. \nonumber
        \end{aligned}
    \end{equation}
where $l_{\mathcal{C}}^{max}\eqqcolon\max_{\boldsymbol{x}_t\in\mathcal{D}}\parallel \boldsymbol{x}_t - \mathcal{C} \boldsymbol{\psi}(\boldsymbol{x}_t,\boldsymbol{\theta}) \parallel$. Then, the norm of the estimation error $\boldsymbol{e}_t$ is zero, i.e. $\parallel \boldsymbol{e}_t\parallel=0$, if the following conditions are satisfied:
\begin{enumerate}
\item There are infinite number of hidden layer nodes $n_h$.
\item The observation interval $\Delta t$ is sufficiently small such that $\mu_x$ and $\mu_u$ equal to zero.
\item $l_{\mathcal{C}}^{max}$ is zero.
\end{enumerate}
\end{lemma}

\begin{lemma} \label{lemma:dfdx}
If $\parallel \boldsymbol{e}_t\parallel=0$, the partial derivative of the estimated dynamics $\hat{\boldsymbol{f}}$ with respect to states and inputs, denoted as $\frac{\partial\hat{\boldsymbol{f}}}{\partial\boldsymbol{x}_t}$ and $\frac{\partial\hat{\boldsymbol{f}}}{\partial\boldsymbol{u}_t}$, are equal to the true partial derivatives $\frac{\partial\boldsymbol{f}}{\partial\boldsymbol{x}_t}$ and $\frac{\partial\boldsymbol{f}}{\partial\boldsymbol{u}_t}$.
\end{lemma}

\begin{proof}

    The estimated and true partial derivatives, $\frac{\partial\hat{\boldsymbol{f}}}{\partial\boldsymbol{x}_t}$ and $\frac{\partial\boldsymbol{f}}{\partial\boldsymbol{x}_t}$, can be expanded as:
    \begin{equation}
    \begin{aligned}
        \frac{\partial\hat{\boldsymbol{f}}}{\partial\boldsymbol{x}_t}&=\lim_{\Delta\boldsymbol{x}\rightarrow 0}
        \frac{\hat{\boldsymbol{f}}(\boldsymbol{x}_{t+1},\boldsymbol{u}_t)-\hat{\boldsymbol{f}}(\boldsymbol{x}_{t},\boldsymbol{u}_t)}{\Delta\boldsymbol{x}},\\
        \frac{\partial{\boldsymbol{f}}}{\partial\boldsymbol{x}_t}&=\lim_{\Delta\boldsymbol{x}\rightarrow 0}
        \frac{{\boldsymbol{f}}(\boldsymbol{x}_{t+1},\boldsymbol{u}_t)-{\boldsymbol{f}}(\boldsymbol{x}_{t},\boldsymbol{u}_t)}{\Delta\boldsymbol{x}}\nonumber.
    \end{aligned}
    \end{equation}
    where $\Delta\boldsymbol{x}=\boldsymbol{x}_{t+1}-\boldsymbol{x}_{t}$.
    Denote the difference between estimated and true partial derivatives as:
    \begin{equation}
    \begin{aligned}
    &\Delta \frac{\partial\boldsymbol{f}}{\partial\boldsymbol{x}_t}=\frac{\partial\hat{\boldsymbol{f}}}{\partial\boldsymbol{x}_t}-\frac{\partial{\boldsymbol{f}}}{\partial\boldsymbol{x}_t}\\
    &={\footnotesize\lim_{\Delta\boldsymbol{x}\rightarrow 0}\frac{(\hat{\boldsymbol{f}}(\boldsymbol{x}_{t+1},\boldsymbol{u}_t)-\boldsymbol{f}(\boldsymbol{x}_{t+1},\boldsymbol{u}_t))-(\hat{\boldsymbol{f}}(\boldsymbol{x}_{t},\boldsymbol{u}_t)-{\boldsymbol{f}}(\boldsymbol{x}_{t},\boldsymbol{u}_t))}{\Delta\boldsymbol{x}}\nonumber}.
    \end{aligned}
    \end{equation}
If $\parallel \boldsymbol{e}_t\parallel=0$, 
\begin{equation}
    \begin{aligned}
    \hat{\boldsymbol{f}}(\boldsymbol{x}_{t+1},\boldsymbol{u}_t)-\boldsymbol{f}(\boldsymbol{x}_{t+1},\boldsymbol{u}_t)&=0,\\
    \hat{\boldsymbol{f}}(\boldsymbol{x}_{t},\boldsymbol{u}_t)-{\boldsymbol{f}}(\boldsymbol{x}_{t},\boldsymbol{u}_t)&=0.
    \end{aligned}
\end{equation}
Therefore,
\begin{equation}
    \lim_{\parallel \boldsymbol{e}_t\parallel\rightarrow 0}\Delta \frac{\partial\boldsymbol{f}}{\partial\boldsymbol{x}_t}= 0.
\end{equation}
Similar proof also leads to:
\begin{equation}
    \lim_{\parallel \boldsymbol{e}_t\parallel\rightarrow 0}\Delta \frac{\partial\boldsymbol{f}}{\partial\boldsymbol{u}_t}= 0,
\end{equation}
where $\Delta \frac{\partial\boldsymbol{f}}{\partial\boldsymbol{u}_t}=\frac{\partial\hat{\boldsymbol{f}}}{\partial\boldsymbol{u}_t}-\frac{\partial{\boldsymbol{f}}}{\partial\boldsymbol{u}_t}$.
\end{proof}

\begin{proof}[\textbf{Proof of Theorem \ref{theorem}}]

Suppose we know the true dynamics $\boldsymbol{f}$, one can generate a matrix $\boldsymbol{F}$ with data $\boldsymbol{\zeta}$ as in \eqref{eq:Fw=0}, where $\boldsymbol{F}=\hat{\boldsymbol{F}}-\Delta\boldsymbol{F}$, with
\begin{equation}
    \Delta\boldsymbol{F} =
    \begin{bmatrix}
        \Delta\boldsymbol{A}&\mathbf{0}&\Delta\boldsymbol{V}\\
\Delta\boldsymbol{B}&\mathbf{0}&\mathbf{0}
    \end{bmatrix}
    .
\end{equation}
where
\begin{align}
\Delta\boldsymbol{A}&=
\begin{bmatrix}
0 & \Delta\frac{-\partial \boldsymbol{f}^\prime}{\partial \boldsymbol{x}_{\ubar{t}+1}} & \cdots  & 0 & 0\\
0   & 0 & \cdots & 0  & 0\\
\vdots   & \vdots & \ddots &  \ddots & \vdots \\
0 & 0  &  \cdots & 0 & \Delta\frac{-\partial \boldsymbol{f}^\prime}{\partial \boldsymbol{x}_{\bar{t}-1}} \\
0 & 0 & \cdots & 0 & 0    
\end{bmatrix}
,\\
\Delta\boldsymbol{B}&=
\begin{bmatrix}
 \Delta\frac{\partial \boldsymbol{f}^\prime}{\partial \boldsymbol{u}_{\ubar{t}}} & 0  & \cdots  & 0 \\
0 & \Delta\frac{\partial \boldsymbol{f}^\prime}{\partial \boldsymbol{u}_{\ubar{t}+1}} &  \cdots  & 0 \\
 \vdots &  &  \ddots & \vdots  \\
 0 & 0  & \cdots &\Delta\frac{\partial \boldsymbol{f}^\prime}{\partial \boldsymbol{u}_{\bar{t}-1}}
\end{bmatrix}
,\\
\Delta\boldsymbol{V}&=
\begin{bmatrix}
0 & 0  & \cdots & 0 & \Delta\frac{\partial \boldsymbol{f}}{\partial \boldsymbol{x}_{\bar{t}}}
\end{bmatrix}^{\prime}
.
\end{align}
Equation \eqref{eq:Fw=0} can be written as:
\begin{equation}
    0=\hat{\boldsymbol{F}}\hat{\boldsymbol{\nu}}=(\boldsymbol{F}+\Delta\boldsymbol{F})\hat{\boldsymbol{\nu}}=\boldsymbol{F}\hat{\boldsymbol{\nu}}+\Delta\boldsymbol{F}\hat{\boldsymbol{\nu}}.
\end{equation}
According to Lemma \ref{lemma:estimationError}, the norm of the estimation error $\boldsymbol{e}_t$ is zero if 
\begin{enumerate}
    \item There is an infinite number of hidden layer nodes in DNN.
    \item The observation interval $\Delta t$ is sufficiently small, which makes $\mu_x$ and $\mu_u$ equal to zero.
    \item $l_{\mathcal{C}}^{max}$ equals to zero.
\end{enumerate}
Once the above conditions are fulfilled, we then employ Lemma \ref{lemma:dfdx}. According to Lemma \ref{lemma:dfdx}, if the norm of the estimation error is zero, $\Delta\frac{\partial\boldsymbol{f}}{\partial\boldsymbol{x}}$ and $\Delta\frac{\partial\boldsymbol{f}}{\partial\boldsymbol{u}}$ are zero. Therefore, $\Delta \boldsymbol{A},\Delta \boldsymbol{B},\Delta \boldsymbol{V}$ become zero matrices, which means $\Delta \boldsymbol{F}$ becomes a zero matrix. As a result, $\boldsymbol{F}=\hat{\boldsymbol{F}}$, $\hat{\boldsymbol{\nu}}$ converges to $\boldsymbol{\nu}$ which is obtained using the true $\boldsymbol{F}$. Thus, $\hat{\boldsymbol{\omega}}$ converges to $\boldsymbol{\omega}$.
\end{proof}

\section{Numerical Experiments}

The proposed data-driven inverse optimal control algorithm is evaluated by a simulation of a pendulum model. 

\begin{figure}[H]
    \centering
    \includegraphics[width=0.2\textwidth]{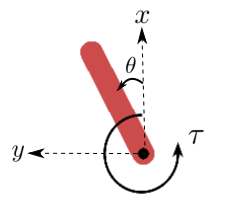}
    \caption{A simulated pendulum.}
    \label{fig:Pendulum}
\end{figure}

As shown in Figure \ref{fig:Pendulum}, we consider that a pendulum system moves in the vertical plane with continuous dynamics given by \cite{han2020deep}
\begin{equation} \label{eq:pendulum}
ml^2\ddot\theta + mgl\sin\theta = \tau,
\end{equation}
where $\theta$ is the angle of the pendulum, $m$ is the mass of the pendulum, $g$ is the gravitational acceleration, $l$ is the length of the pendulum and $\tau$ here is the torque applied. The parameters used are $g=10m/s^2$, the length $l=10m$, and the pendulum mass $m=1kg$. By defining the states and control inputs of the pendulum:
\begin{equation}
    \boldsymbol x\triangleq\begin{bmatrix}
\theta&\dot{\theta}
\end{bmatrix}^{\prime}\quad \text{and} \quad \boldsymbol u\triangleq\boldsymbol\tau,
\end{equation}
respectively, one could write \eqref{eq:pendulum} in state-space representation  $\boldsymbol{\dot x}=\boldsymbol g(\boldsymbol x, \boldsymbol u)$  and further approximate it by the following discrete-time  form
\begin{equation}\label{equ_discretization}
\boldsymbol x_{t+1}\approx\boldsymbol x_t+\Delta\cdot\boldsymbol g(\boldsymbol x_t, \boldsymbol u_{t})\triangleq\boldsymbol{f}(\boldsymbol x_t, \boldsymbol u_{t}),
\end{equation}
where $\Delta =0.001 \mathrm{s}$ is the discretization interval. The motion of the pendulum is controlled to minimize the objective function \eqref{eq:objective_unknown}, which here is set as a weighted distance to the goal state $\boldsymbol{x}^{\text{g}}=[{\theta}^\text{g}, {\dot{\theta}}^\text{g}]^\prime=[\pi,0]^\prime$ plus the control effort $\norm{\boldsymbol{u}}^2$. Here, the corresponding  features and weights defined are as follows.
\begin{align}
\label{eq:PENOCObj}
    \boldsymbol{\phi}&= 
      \begin{bmatrix}
      ({\theta}-{\theta}^\text{g})^2\\
      ({{\dot\theta}}-{{\dot\theta}}^\text{g})^2\\
      ||\boldsymbol{u}||^2
    \end{bmatrix}, \qquad
    \boldsymbol{\omega}=\begin{bmatrix}
    2 \\ 1 \\ 1 
    \end{bmatrix}, 
\end{align}

The initial condition of the robot arm is set as $ x_0=[0,0]^{\prime}$, and the time horizon is set as $T=10$. We set the ground-truth weights as in (\ref{eq:PENOCObj}). 

In the data-driven IOC task, we learn the weight vector $\boldsymbol{\omega}$ and the system $\boldsymbol{f}$ from the segment data of the optimal trajectory. The effectiveness of the algorithm is demonstrated in Table~\ref{table:minicost} and \ref{table:nh}. We tested our algorithm with different results of $l_{\mathcal{C}}^{max}$ and different $n_h$. In our case, we consider the observation interval $\Delta t$ to be the same as the sampling instance interval of the unknown discrete-time system \eqref{eq:dyn_unknown}. Therefore, the constants $\mu_x$ and $\mu_u$ are kept unchanged.

\begin{table}[h] 
\centering
	\caption	{IOC results with different $l_{\mathcal{C}}^{max}$.}
	\begin{tabular}{llll}
		\toprule
		$l_{\mathcal{C}}^{max}$  &   $\boldsymbol{\hat{\omega}}$ & Error, Weight &  Error, Traj. \\
		\midrule
		1e-4 &    $[1.75, 1.37, 0.87]$               & $0.468$ &6.13e-2 \\[10pt]
		1e-5 & $[1.89, 1.16, 0.94]$                         & $0.207$ &1.25e-3 \\[10pt]
		1e-6 & $[1.92, 1.12, 0.96]$                          & $0.146$ &5.22e-4 \\[10pt]
		
		\bottomrule
	\end{tabular}
	\label{table:minicost}
\end{table}

\begin{table}[h] 
\centering
	\caption	{IOC results with different number of hidden layer nodes $n_h$ in DNN .}
	\begin{tabular}{llll}
		\toprule
		$n_h$  &   $\boldsymbol{\hat{\omega}}$ & Error, Weight &  Error, Traj.    \\
		\midrule
		1254 &    $[1.71, 1.24, 1.05]$               & $0.382$  &6.83e-2\\[10pt]
		1670 & $[1.93, 1.15, 0.92]$                         & $0.183$ & 1.70e-2\\[10pt]
		2086 & $[1.96, 1.05, 0.98]$                          & $0.070$ & 2.24e-4\\[10pt]
		
		\bottomrule
	\end{tabular}
	\label{table:nh}
\end{table}

In Table~\ref{table:minicost} and Table~\ref{table:nh}, it is shown that as $l_{\mathcal{C}}^{max}$ decreases, or the number of hidden layer nodes $n_h$ increases, the IOC result gets closer to the ground truth, i.e. the 2-norm error between $\hat{\boldsymbol{\omega}}$ and the true $\boldsymbol{\omega}$ decreases. Optimal trajectories are also generated using the estimate $\hat{\boldsymbol{\omega}}$. We can see that as the error of weight decreases, the 2-norm of the error between the ground truth trajectory and trajectories produced using $\hat{\boldsymbol{\omega}}$ decreases.

\section{Conclusions}
This paper has developed a data-driven method to solve the
inverse optimal control problem with unknown system dynamics. The unknown system dynamics is approximated by the finite-dimensional Koopman operator, with the observables represented by a deep neural network. We proved that if certain conditions are fulfilled, the approximated system, as well as the approximated system derivatives with respect to states and inputs, will converge to the true system and the true derivatives. As a result, an iterative scheme to update the least-square estimate of the weight vector in the unknown system dynamics is proposed. 

For future research, we will extend the proposed method to an IOC method with noisy data. The motivation here is that the assumption of perfect data is sometimes challenging to fulfill since there are always errors when obtaining data using sensors in reality. Another potential research direction would be performing IOC with system output rather than the states.

\bibliographystyle{ieeetr}

\bibliography{trobib,Shaoshuai}           

\end{document}